\documentclass[letter]{aa} % for the letters 
\usepackage{natbib}
\usepackage{graphicx}
\usepackage{txfonts}
\begin{document}
   \title{A scenario of planet erosion by coronal radiation}

   \author{J. Sanz-Forcada\inst{1}
          \and
          I. Ribas\inst{2}
          \and
          G. Micela\inst{3}
          \and
          A. M. T. Pollock\inst{4}
          \and
          D. Garc\'{i}a-\'Alvarez\inst{5,6}
          \and
          E. Solano\inst{1,7}
          \and
          C. Eiroa\inst{8}
          }

   %\offprints{J. Sanz-Forcada, \email{jsanz@cab.inta-csic.es}}

   \institute{Laboratorio de Astrof\'{i}sica Estelar y Exoplanetas,
     Centro de Astrobiolog\'{i}a / CSIC-INTA, LAEFF Campus, P.O. Box 78, 
     E-28691 Villanueva de la Ca\~nada, Madrid, Spain; %\\
     \email{jsanz@cab.inta-csic.es}
         \and
     Institut de Ci\`ences de l'Espai (CSIC-IEEC), Campus UAB, Fac.
     de Ci\`encies, Torre C5-parell-2$^{\underline a}$ planta, 
     E-08193 Bellaterra, Spain%\\
%     \email{iribas@ieec.uab.es}
     \and
     INAF -- Osservatorio Astronomico di Palermo
     G. S. Vaiana, Piazza del Parlamento, 1; Palermo, I-90134, Italy%\\
%     \email{giusi@astropa.inaf.it}
     \and
     XMM-Newton SOC, European Space Agency, ESAC, Apartado 78,
     E-28691 Villanueva de la Ca\~nada, Madrid, Spain%\\ 
%     \email{Andy.Pollock@esa.int}
     \and
     Instituto de Astrof\'{i}sica de Canarias, E-38205 La Laguna, Spain %\\
     \and
     Grantecan CALP, E-38712 Bre\~na Baja, La Palma, Spain%\\
     \and
     Spanish Virtual Observatory, Centro de Astrobiolog\'{i}a /
     CSIC-INTA, LAEFF Campus, Madrid, Spain%\\
%    \email{esolano@cab.inta-csic.es}
     \and
     Dpto. de F\'{i}sica Te\'orica, C-XI, Facultad de Ciencias, 
     Universidad Aut\'onoma de Madrid, Cantoblanco, E-28049 Madrid, Spain%\\
%    \email{carlos.eiroa@uam.es})
             }

   \date{Received 14 November 2009; accepted 8 February 2010}

% \abstract{}{}{}{}{} 
% 5 {} token are mandatory
 
  \abstract
   {According to theory, high-energy emission from the coronae of cool
     stars can severely erode the atmospheres of orbiting planets. No
     observational tests of the long-term erosion effects have been 
     made yet.}
   {We analyze the current distribution of planetary mass with X-ray
     irradiation of the atmospheres to make an observational
     assessment of the consequences of erosion by coronal radiation.}
   {We studied a large sample of planet-hosting stars with XMM-Newton,
     Chandra, and ROSAT, 
     carefully identified the X-ray counterparts,
     and fit their spectra to
     accurately measure the stellar X-ray flux.}
   {The distribution of the planetary masses with X-ray flux
     suggests that erosion has taken place. Most surviving
     massive planets 
     ($M_{\rm p} \sin i >1.5$\,M$_{\rm J}$) have been exposed to lower
     accumulated irradiation. Heavy erosion during the
     initial stages of stellar evolution is followed by a phase of 
     much weaker erosion. A line dividing these two phases could be
     present, showing a strong dependence on planet 
     mass. Although a larger sample will be required to establish
     a well-defined erosion line, the distribution found is
     very suggestive.} 
   {The distribution of planetary mass with X-ray flux is consistent
     with a scenario in which planet atmospheres have suffered the
     effects of erosion by coronal X-ray and EUV emission. 
     The erosion line is
     an observational constraint for models of atmospheric erosion.} 

   \maketitle
%
%________________________________________________________________

%vvvvvvvvvvvvvvvvvvvvvvvvvvvvvvvvvvvvvvvvvvvvvvvvvvvvv
\section{Introduction}

The expected effects on the erosion of exoplanetary atmospheres by stellar radiation
have been the subject of much theoretical work
\citep{lam03,bar04,yel04,rib05,tia05,cec06,lec07,gar07,erk07,hub07,pen08a,pen08b,cec09,dav09}. 
In planets around late-type stars, stellar radiation in the
X-ray ($\sim$1--100~\AA)
and
EUV ($\sim$100--900 \AA)
ranges has the strongest effect on atmospheric evaporation.  
Late-type stars
are copious emitters of X-ray and EUV radiation from
high-temperature ($\sim$1--30~MK) material in coronae,
whose development is favoured by fast rotation making it
most important for young stars that retain much of the angular
momentum of the parent cloud. Observations of
stellar clusters have shown that X-ray emission decreases with
age, as the rotation slows \citep[c.f.][]{fav03}.
The sample of known exoplanets is now large enough for
an observational search to be made for the erosion effects 
on planet masses. 

After planet formation and once the protoplanetary disk has dissipated,
a planet is exposed to high levels of 
coronal emission from the rapidly rotating young host star,
which is much stronger for
closer-in planets. This emission is expected to
progressively erode the planet atmosphere through
evaporation (thermal losses)
mediated by gravity.
A planetary magnetic field should provide some protection against
losses of ionized material,
although little work
has been done on these effects \citep[e.g.][and references therein]{gri09}.
Once the star slows down and becomes more
X-ray and EUV quiet, the planet mass decreases at a lower rate. 
The relatively simple approach proposed by \citet{wat81} and
subsequently modified by \citet{lam03}, \citet{bar04}, and
\citet{erk07} to account for the expansion radius of the atmosphere
$R_1$ ($R_1 \ge R_{\rm p}$)
and the filling of the
Roche lobe (using the $K$ parameter, with $K \le 1$)
leads to the following expression for the thermal planetary mass-loss rate
\begin{equation}
 \dot M=\frac{4 \pi \beta^3 R_{\rm p}^3 F_{\rm XUV}}{{\rm G} K M_{\rm  p}}
\end{equation}
where $\beta = R_1/R_{\rm p}$,
$F_{\rm XUV}$ is the X-ray and EUV flux at the planet orbit, 
and G is the gravitational constant.

We have adopted the direct experimental approach of
examining the dependence of planet mass on the X-ray
flux it has received. 
{\em In the long term} the effects of erosion by atmospheric losses should
result in an uneven distribution 
of planet masses with the X-ray flux at the planet orbit. 
We set up a database of X-ray and EUV emission of the stars
hosting exoplanets \citep[``X-exoplanets'',][]{san09} to facilitate
analysis of the effects of coronal radiation on exoplanet
atmospheres. In this work we present the results of a study with the whole
sample of 59 non-giant stars hosting 75 exoplanets with masses up to 
8\,M$_J$ that have been observed in X-rays with XMM-Newton, Chandra, or
ROSAT,
making what we consider to be the safe assumption that
the X-ray flux is proportional to the entire evaporating flux,
because both X-rays and EUV
stem from similar processes in the corona of the star.
Present-day telescopes give no access to stellar EUV observations,
which would also be limited by absorption in the insterstellar medium. 
In Sect.~2 we describe the observations, data
reduction, and results, before discussing their
implications in Sect. 3, and finish with the
conclusions in Sect.~4.

%----------------------------------  Fig. 1
   \begin{figure}[t]
   \centering
   \includegraphics[width=0.49\textwidth,clip]{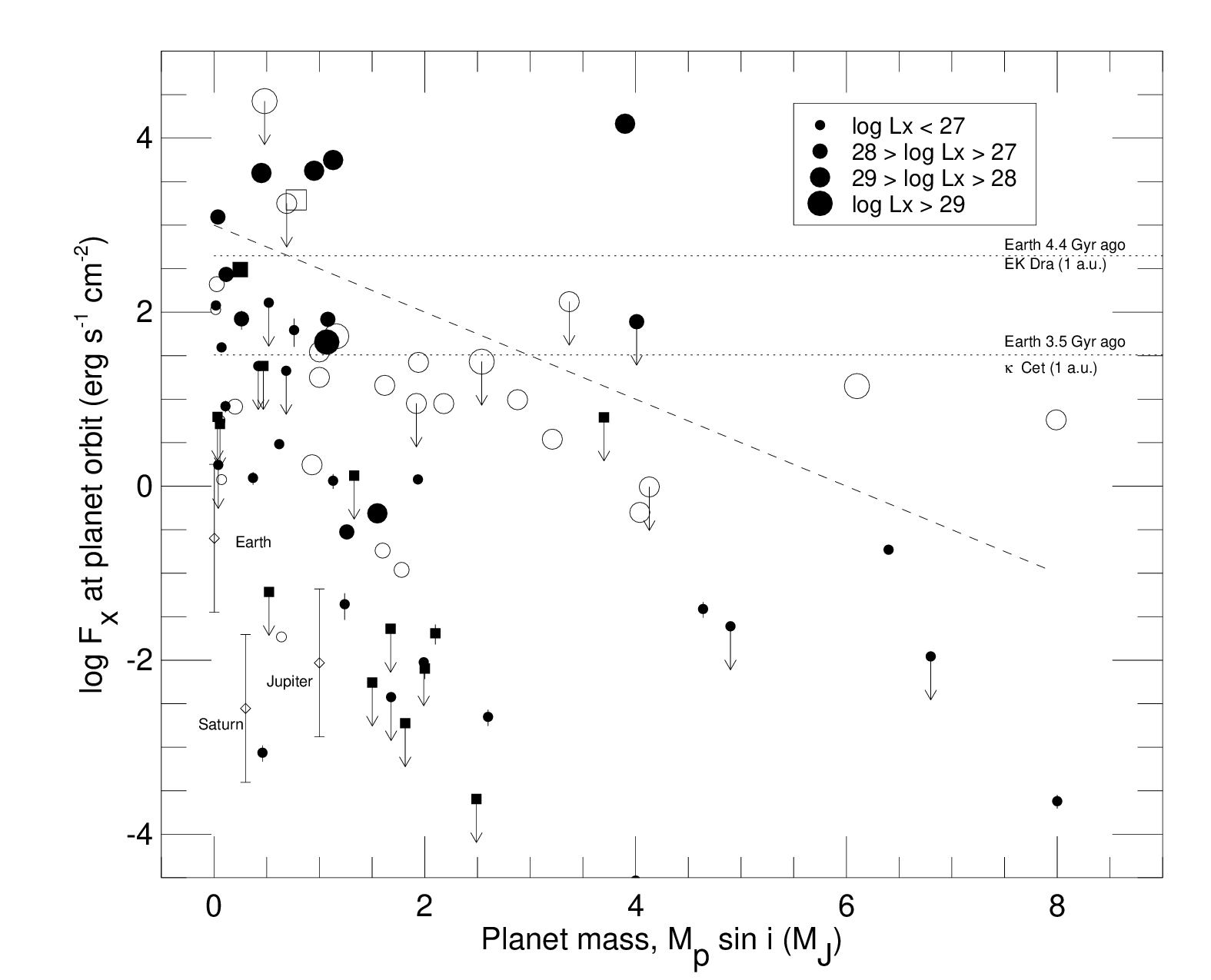}
   \caption{Distribution of planetary masses ($M_{\rm p} \sin i$) with X-ray
     flux at the planet orbit. Filled symbols (squares for subgiants,
     circles for dwarfs) are
     XMM-Newton and Chandra data. Arrows indicate upper
     limits. Open symbols are ROSAT data without error
     bars. Diamonds represent Jupiter, Saturn, and the Earth. 
     The dashed line marks the ``erosion line''. 
     Dotted lines indicate the X-ray flux of the younger Sun
     at 1 a.u.}\label{masses}
    \end{figure}
%----------------------------------------------

%vvvvvvvvvvvvvvvvvvvvvvvvvvvvvvvvvvvvvvvvvvvvvvvvvvvvv
\section{Observations and results}\label{sec:results}
The XMM-Newton and Chandra X-ray observations shown in
Table~\ref{tabfluxes}
were reduced following standard procedures following corrections for
proper motion.
These corrections were
particularly important in some cases for discarding
erroneous detections reported in the literature.
For 47 UMa, for example,
we find $\log L_{\rm X}\sim25.45$~erg\,s$^{-1}$
compared to the value of $\log L_{\rm X}=27.13$~erg\,s$^{-1}$
estimated by \citet{kas08}.
The star was
detected near its expected position $\alpha$=10:59:28.0,
$\delta$=40:25:46 close to a brighter source
$\alpha$=10:59:26.7, $\delta$=40:26:04: the instrumental spatial
  resolution is 6\arcsec.
Tentative evidence for detection of a Fe K$\alpha$ emission line
suggests the brighter source may be a highly obscured AGN at z$\sim0.2$
(G. Miniutti, private comm.). Similarly for HD~209458, 
we calculated an upper limit of $\log L_{\rm X}<26.12$~erg\,s$^{-1}$,
a value much lower than reported elsewhere by
\cite{kas08} and \cite{pen08a}, who might 
have confused the star with a nearby object. 
 
%----------------------------------  Fig. 2
   \begin{figure}[t]
   \centering
   \includegraphics[width=0.49\textwidth,clip]{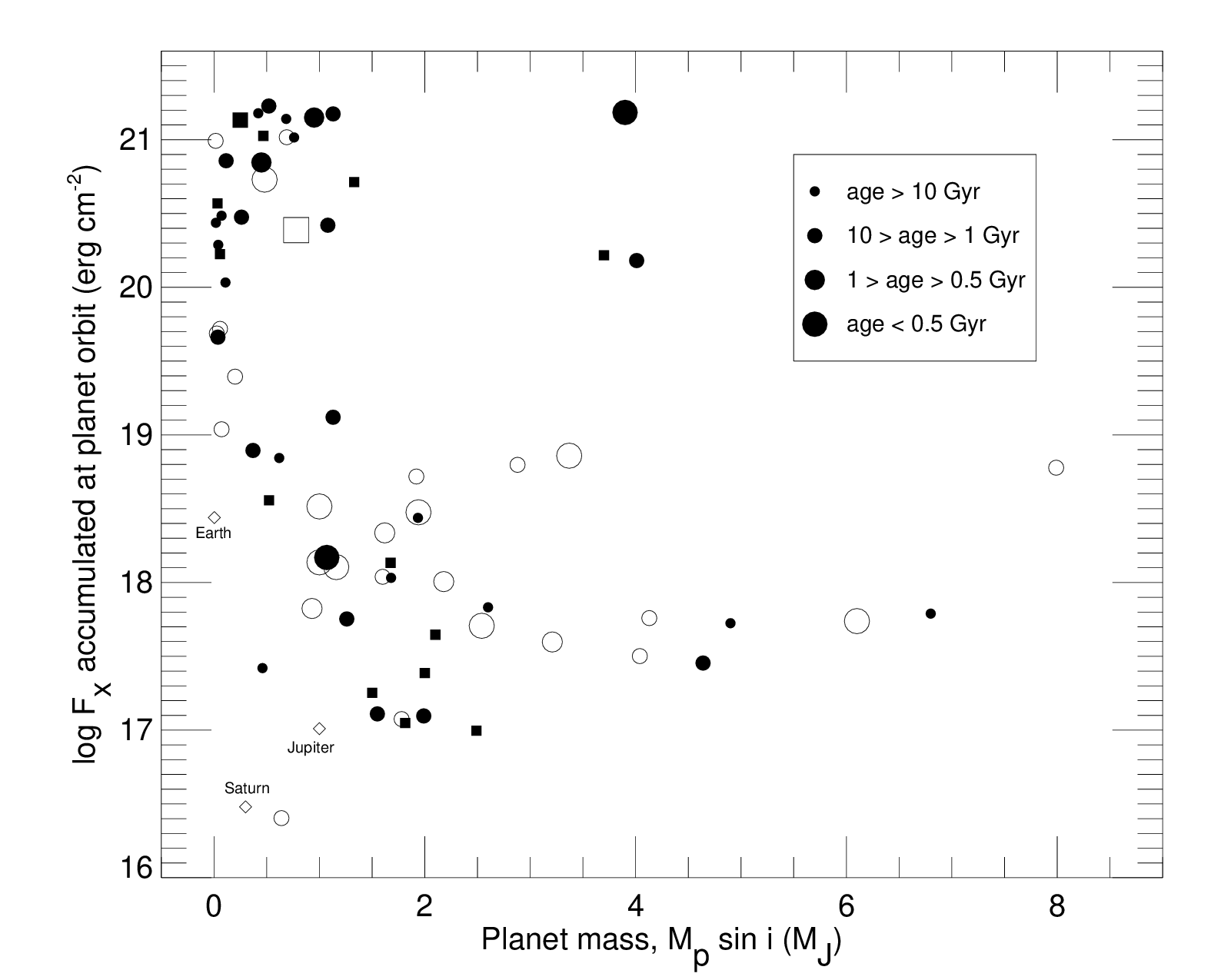}
   \caption{Distribution of planetary masses ($M_{\rm p} \sin i$) with the X-ray
     flux accumulated at the planet orbit since an age of 20 Myrs to the
     present day (see text). Symbols as in Fig.~\ref{masses}.}\label{agemasses}
    \end{figure}
%----------------------------------------------

Extracted spectra were fit using standard procedures with
coronal models of 1 to 3 temperature components \citep[see
  e.g.,][]{san03a}.  The actual 
model used in the fit has little influence on the
calculated X-ray (0.12--2.48~keV or 5--103~\AA) flux shown in
Table~\ref{tabfluxes}. More  
details on the data reduction and treatment will be given in Sanz-Forcada et
al. 2010 (in prep.). Measurements with S/N$<$3 were
considered as upper limits. 
We complemented the sample with lower spatial resolution ROSAT
measurements, excluding detections with low statistics
(S/N$<$3), and further marking as upper limits the objects with suspected
X-ray bright companions. The sample of 75 exoplanets including
XMM-Newton, Chandra, and ROSAT detections, have been
compared with the whole exoplanet database (417 objects to date) 
to check whether our
sample is representative of the known exoplanets. We applied the
Kolmogorov-Smirnov test to compare both samples: and they represent the
same distributions in mass (99.7\% probability) or period (92.8\%
probability) in single variable tests.

%----------------------------------  Table 1
\onltab{1}{
\begin{table*}[t]
\caption[]{X-ray flux ($0.12-2.48$~keV) of stars with exoplanets$^a$}\label{tabfluxes}
\tabcolsep 3.pt
\begin{center}
\begin{scriptsize}
  \begin{tabular}{lcrrcccccccccclr}
\hline \hline
{Planet name} & {Sp. type} & \multicolumn{2}{c}{Measured coordinates} 
& {Stellar distance} & {$\log L_{\rm X}$}
& {S/N} & age & $M_p \sin i$ & $a_p$ & $\log F_{\rm X}$ & 
$\log F_{\rm X}$accum. & $\rho \dot M_{\rm X}$ & {Instr.$^b$} &  {Date} & {t} \\ 
{} & {(star)} & \multicolumn{2}{c}{$\alpha$, $\delta$ (J2000.0)} & {(pc)} & {(erg s$^{-1}$)} & {($L_{\rm X}$)} & (Gyr) &
{(m$_J$)} & {(a.u.)} & 
\multicolumn{2}{l}{(erg s$^{-1}$cm$^{-2}$) \, (erg cm$^{-2}$)} & {(g$^2$s$^{-1}$cm$^{-3}$)$^c$} 
& &   & (ks)\\ 
\hline
%--------------------------------------------------------------
14 Her b            &        K0V & 16:10:24.6 & $+43$:49:01 & $ 18.10\pm  0.19$ &     26.92 &    4.9 &    7.47  &  4.64 &   2.77 & $ -1.41 $ & 17.45 &      1.7e+06 & EPIC & 2005/09/11 &   5 \\
16 Cyg B b          &      G2.5V & 19:41:48.9 & $+50$:31:28 & $ 21.41\pm  0.23$ &  $<$25.48 &    1.7 &  $ <15$  &  1.68 &   1.68 & $<-2.43 $ & 18.03 &    (1.7e+05) & EPIC & 2008/11/08 &  11 \\
2M1207 b            &         M8 & 12:07:33.5 & $-39$:32:54 & $ 52.40\pm  1.10$ &  $<$26.24 &    0.4 &  $ <15$  &  4.00 &  46.00 & $<-4.53 $ & 12.85:&    (1.3e+03) & ACIS & 2003/03/03 &  50 \\
47 UMa b            &        G0V & 10:59:28.4 & $+40$:25:46 & $ 13.97\pm  0.13$ &     25.45 &    4.8 &  $ <15$  &  2.60 &   2.11 & $ -2.65 $ & 17.83 &      1.0e+05 & EPIC & 2006/06/11 &   8 \\
47 UMa c            &            &            &             &                   &           &        &          &  0.46 &   3.39 & $ -3.07$ & 17.42  &      3.9e+04 & & & \\
51 Peg b            &       G2IV & 22:57:28.1 & $+20$:46:08 & $ 15.36\pm  0.18$ &  $<$26.26 &    2.6 &  $ <15$  &  0.47 &   0.05 & $< 1.38$ & 21.03: &    (1.1e+09) & ACIS & 2008/12/06 &   5 \\
$\beta$ Pic b       &        A6V & 05:47:17.1 & $-51$:03:59 & $ 19.30\pm  0.19$ &     25.63 &    5.7 &  $ <15$  &  8.00 &   8.00 & $ -3.62 $ & 16.94 &      1.1e+04 & EPIC & 2004/01/04 &  68 \\
$\epsilon$ Eri b&        K2V & 03:32:55.9 & $-09$:27:31 & $  3.20\pm  0.01$ &     28.20 &  296.7 &    1.12  &  1.55 &   3.39 & $ -0.31 $ & 17.11 &      2.2e+07 & EPIC & 2003/01/19 &  12 \\
GJ 436 b            &       M2.5 & 11:42:11.6 & $+26$:42:16 & $ 10.20\pm  0.24$ &     25.96 &   14.5 &  $ <15$  &  0.07 &   0.03 & $  1.60 $ & 20.48 &      1.8e+09 & EPIC & 2008/12/10 &  30 \\
GJ 674 b            &       M2.5 & 17:28:40.3 & $-46$:53:50 & $  4.54\pm  0.03$ &     27.73 &  178.5 &    2.84  &  0.04 &   0.04 & $  3.09 $ & 19.66 &      5.6e+10 & EPIC & 2008/09/05 &  44 \\
GJ 86 b             &        K1V & 02:10:28.1 & $-50$:49:19 & $ 11.00\pm  0.07$ &  $<$27.42 &   43.0 &    3.56  &  4.01 &   0.11 & $< 1.89 $ & 20.18 &    (3.5e+09) & EPIC & 2008/06/10 &  15 \\
GJ 876 b            &        M4V & 22:53:17.3 & $-14$:15:55 & $  4.72\pm  0.05$ &     26.16 &   34.7 &  $ <15$  &  1.93 &   0.21 & $  0.08 $ & 18.44 &      5.4e+07 & EPIC & 2008/11/14 &  23 \\
GJ 876 c            &            &            &             &                   &           &        &          &  0.62 &   0.13 & $  0.48$ & 18.84  &      1.4e+08 & & & \\
GJ 876 d            &            &            &             &                   &           &        &          &  0.02 &   0.02 & $  2.07$ & 20.44  &      5.3e+09 & & & \\
HD 4308 b           &        G5V & 00:44:39.4 & $-65$:39:05 & $ 21.90\pm  0.27$ &  $<$25.84 &    2.5 &  $ <15$  &  0.04 &   0.12 & $< 0.24 $ & 20.29 &    (7.9e+07) & EPIC & 2008/12/02 &   9 \\
HD 20367 b          &         G0 & 03:17:40.1 & $+31$:07:37 & $ 27.00\pm  0.79$ &     29.30 &  139.6 &    0.22  &  1.07 &   1.25 & $  1.65 $ & 18.17 &      2.0e+09 & EPIC & 2005/02/11 &  10 \\
HD 46375 b          &       K1IV & 06:33:12.4 & $+05$:27:49 & $ 33.40\pm  1.19$ &     27.16 &    7.3 &    5.23: &  0.25 &   0.04 & $  2.49$ & 21.13: &      1.4e+10 & EPIC & 2005/10/14 &   8 \\
HD 49674 b          &        G5V & 06:51:30.9 & $+40$:52:03 & $ 40.70\pm  1.89$ &     27.41 &    6.5 &    3.62  &  0.12 &   0.06 & $  2.43 $ & 20.86 &      1.2e+10 & EPIC & 2006/04/10 &   8 \\
HD 50554 b          &         F8 & 06:54:42.8 & $+24$:14:43 & $ 31.03\pm  0.97$ &  $<$26.59 &    2.7 &   12.21  &  4.90 &   2.38 & $<-1.61 $ & 17.72 &    (1.1e+06) & EPIC & 2006/04/16 &   9 \\
HD 52265 b          &        G0V & 07:00:18.0 & $-05$:22:01 & $ 28.00\pm  0.66$ &     26.89 &    5.4 &    7.82  &  1.13 &   0.49 & $  0.06 $ & 19.12 &      5.2e+07 & EPIC & 2008/09/19 &   9 \\
HD 70642 b          &     G5IV-V & 08:21:28.2 & $-39$:42:18 & $ 29.00\pm  0.50$ &     26.39 &    4.2 &  $ <15$  &  2.00 &   3.30 & $ -2.10$ & 17.39: &      3.6e+05 & EPIC & 2006/04/08 &  13 \\
HD 75289 b          &        G0V & 08:47:40.1 & $-41$:44:14 & $ 28.94\pm  0.47$ &  $<$26.16 &    2.2 &  $ <15$  &  0.42 &   0.05 & $< 1.38 $ & 21.18 &    (1.1e+09) & EPIC & 2005/04/28 &   8 \\
HD 93083 b          &        K3V & 10:44:20.9 & $-33$:34:38 & $ 28.90\pm  0.84$ &     26.90 &    6.3 &    7.71  &  0.37 &   0.48 & $  0.09 $ & 18.89 &      5.6e+07 & EPIC & 2008/05/26 &  12 \\
HD 99492 b          &        K2V & 11:26:45.9 & $+03$:00:24 & $ 18.00\pm  1.07$ &     26.55 &   11.3 &   13.00  &  0.11 &   0.12 & $  0.92 $ & 20.03 &      3.7e+08 & EPIC & 2008/06/19 &  24 \\
HD 102195 b         &        K0V & 11:45:42.2 & $+02$:49:16 & $ 28.98\pm  0.97$ &     28.43 &   53.1 &    0.80  &  0.45 &   0.05 & $  3.60 $ & 20.85 &      1.8e+11 & EPIC & 2008/06/15 &  18 \\
HD 108147 b         &     F8/G0V & 12:25:46.2 & $-64$:01:20 & $ 38.57\pm  1.03$ &     27.39 &    4.2 &    3.74  &  0.26 &   0.10 & $  1.92 $ & 20.48 &      3.7e+09 & EPIC & 2002/08/10 &   6 \\
HD 111232 b         &        G8V & 12:48:51.8 & $-68$:25:29 & $ 29.00\pm  0.67$ &  $<$26.08 &    0.9 &  $ <15$  &  6.80 &   1.97 & $<-1.96 $ & 17.79 &    (4.9e+05) & EPIC & 2008/07/29 &   9 \\
HD 114386 b         &        K3V & 13:10:39.7 & $-35$:03:20 & $ 28.00\pm  1.04$ &     26.53 &    3.0 &   13.44  &  1.24 &   1.65 & $ -1.36 $ & 17.76 &      2.0e+06 & EPIC & 2008/07/29 &   9 \\
HD 130322 b         &        K0V & 14:47:32.8 & $-00$:16:54 & $ 30.00\pm  1.34$ &     27.26 &    7.7 &    4.55  &  1.08 &   0.09 & $  1.92 $ & 20.42 &      3.7e+09 & EPIC & 2005/07/21 &   7 \\
HD 160691 b         &     G3IV-V & 17:44:08.7 & $-51$:50:04 & $ 15.30\pm  0.19$ &  $<$26.16 &    2.2 &  $ <15$  &  1.68 &   1.50 & $<-1.64$ & 18.13: &    (1.0e+06) & EPIC & 2008/10/02 &  10 \\
HD 160691 c         &            &            &             &                   &           &        &          &  0.03 &   0.09 & $< 0.79$ & 20.57: &    (2.8e+08) & & & \\
HD 160691 d         &            &            &             &                   &           &        &          &  0.52 &   0.92 & $<-1.22$ & 18.56: &    (2.7e+06) & & & \\
HD 160691 e         &            &            &             &                   &           &        &          &  1.81 &   5.24 & $<-2.73$ & 17.05: &    (8.4e+04) & & & \\
HD 179949 b         &        F8V & 19:15:33.3 & $-24$:10:46 & $ 27.00\pm  0.59$ &     28.38 &  100.9 &    0.86  &  0.95 &   0.05 & $  3.62 $ & 21.15 &      1.9e+11 & ACIS & 2005/05/29 & 150 \\
HD 187123 b         &         G5 & 19:46:57.9 & $+34$:25:09 & $ 50.00\pm  1.63$ &  $<$26.80 &    1.4 &    8.91  &  0.52 &   0.04 & $< 2.11 $ & 21.23 &    (5.7e+09) & EPIC & 2006/04/21 &  16 \\
HD 187123 c         &            &            &             &                   &           &        &          &  1.99 &   4.89 & $<-2.03$ & 17.10  &    (4.2e+05) & & & \\
HD 189733 b         &      K1-K2 & 20:00:43.8 & $+22$:42:34 & $ 19.30\pm  0.32$ &     28.18 &   92.5 &    1.15  &  1.13 &   0.03 & $  3.75 $ & 21.17 &      2.5e+11 & EPIC & 2007/04/17 &  43 \\
HD 190360 b         &       G6IV & 20:03:37.9 & $+29$:53:45 & $ 15.89\pm  0.16$ &  $<$26.38 &    1.4 &  $ <15$  &  1.50 &   3.92 & $<-2.26$ & 17.25: &    (2.5e+05) & EPIC & 2005/04/25 &   4 \\
HD 190360 c         &            &            &             &                   &           &        &          &  0.06 &   0.13 & $< 0.71$ & 20.22: &    (2.3e+08) & & & \\
HD 195019 b         &     G3IV-V & 20:28:18.6 & $+18$:46:10 & $ 37.36\pm  1.24$ &  $<$26.52 &    2.7 &   13.50: &  3.70 &   0.14 & $< 0.79$ & 20.22: &    (2.8e+08) & EPIC & 2006/04/24 &  10 \\
HD 209458 b         &        G0V & 22:03:10.8 & $+18$:53:03 & $ 47.00\pm  2.22$ &  $<$26.12 &    1.8 &  $ <15$  &  0.69 &   0.05 & $< 1.32 $ & 21.14 &    (9.5e+08) & EPIC & 2006/11/15 &  31 \\
HD 216435 b         &        G0V & 22:53:38.1 & $-48$:35:55 & $ 33.30\pm  0.81$ &     27.74 &   11.9 &    2.22  &  1.26 &   2.56 & $ -0.53 $ & 17.75 &      1.3e+07 & EPIC & 2006/04/21 &   7 \\
HD 216437 b         &     G4IV-V & 22:54:39.6 & $-70$:04:26 & $ 26.50\pm  0.41$ &     26.62 &    4.0 &   11.69: &  2.10 &   2.70 & $ -1.69$ & 17.65: &      9.1e+05 & EPIC & 2005/04/13 &   6 \\
HD 217107 b         &       G8IV & 22:58:15.7 & $-02$:23:43 & $ 19.72\pm  0.29$ &  $<$25.30 &    2.3 &  $ <15$  &  1.33 &   0.07 & $< 0.12$ & 20.71: &    (5.9e+07) & EPIC & 2005/05/16 &   7 \\
HD 217107 c         &            &            &             &                   &           &        &          &  2.49 &   5.27 & $<-3.60$ & 17.00: &    (1.1e+04) & & & \\
HD 330075 b         &         G5 & 15:49:37.7 & $-49$:57:48 & $ 50.20\pm  3.75$ &     26.51 &    3.1 &   13.80  &  0.76 &   0.04 & $  1.79 $ & 21.02 &      2.8e+09 & EPIC & 2005/08/07 &  16 \\
$\tau$ Boo b        &        F7V & 13:47:15.9 & $+17$:27:22 & $ 15.60\pm  0.17$ &     28.94 &  317.4 &    0.37  &  3.90 &   0.05 & $  4.16 $ & 21.18 &      6.5e+11 & EPIC & 2003/06/24 &  56 \\
VB 10 b             &        M8V & 19:16:57.3 & $+05$:08:49 & $  6.09\pm  0.13$ &     25.83 &   20.4 &  $ <15$  &  6.40 &   0.36 & $ -0.73 $ & 13.88: &     8.3e+06 & EPIC & 2008/04/07 &  28 \\
%--------------------------------------------------------------
\hline
\multicolumn{14}{c}{ROSAT data} & \multicolumn{2}{c}{Notes}\\
\hline
%--------------------------------------------------------------
61 Vir b            &        G5V & 13:18:24.3 & $-18$:18:40 & $  8.52\pm  0.05$ &     26.88 &    5.7 &    7.96  &  0.02 &   0.05 & $  2.03 $ & 20.99 &      4.8e+09 &  PSPC & \multicolumn{2}{l}{}  \\
61 Vir c            &            &            &             &                   &           &        &          &  0.06 &   0.22 & $  0.75$ & 19.72  &      2.5e+08 & & &  \\
61 Vir d            &            &            &             &                   &           &        &          &  0.07 &   0.48 & $  0.07$ & 19.04  &      5.3e+07 & & &  \\
BD-10 3166 b        &        G4V & 10:58:28.8 & $-10$:46:13 & $ 66:$ &  $<$29.20 &    3.1 &    0.25  &  0.48 &   0.05 & $< 4.42 $ & 20.73 &    (1.2e+12) &  PSPC & \multicolumn{2}{l}{+dM5, unc. $d$}  \\
$\gamma$ Cep b   &        K2V & 23:39:20.8 & $+77$:37:56 & $ 13.79\pm  0.10$ &     27.33 &    8.1 &    4.08  &  1.60 &   2.04 & $ -0.74 $ & 18.04 &      8.1e+06 &  PSPC & \multicolumn{2}{l}{}  \\
GJ 832 b            &            & 21:33:34.0 & $-49$:00:32 & $  4.94\pm  0.03$ &     26.78 &    7.3 &    9.24  &  0.64 &   3.40 & $ -1.73 $ & 16.40 &      8.3e+05 &  PSPC & \multicolumn{2}{l}{}  \\
GJ 3021 b           &        G6V & 00:16:12.7 & $-79$:51:04 & $ 17.62\pm  0.16$ &  $<$28.95 &    7.8 &    0.37  &  3.37 &   0.49 & $< 2.12 $ & 18.86 &    (5.9e+09) &  PSPC & \multicolumn{2}{l}{dM4 in field}  \\
HD 3651 b           &        K0V & 00:39:21.8 & $+21$:15:01 & $ 11.00\pm  0.09$ &     27.27 &    4.5 &    4.46  &  0.20 &   0.28 & $  0.91 $ & 19.39 &      3.7e+08 &  PSPC & \multicolumn{2}{l}{}  \\
HD 10647 b          &        F8V & 01:42:29.3 & $-53$:44:27 & $ 17.30\pm  0.19$ &     28.31 &    N/A &    0.95  &  0.93 &   2.03 & $  0.24 $ & 17.82 &      7.9e+07 &  PSPC & \multicolumn{2}{l}{}  \\
HD 38529 b          &       G4IV & 05:46:34.9 & $+01$:10:05 & $ 42.43\pm  1.66$ &     28.96 &    5.3 &    0.36: &  0.78 &   0.13 & $  3.29$ & 20.39: &      8.7e+10 &  PSPC & \multicolumn{2}{l}{}  \\
HD 41004 A b        &        K1V & 05:59:49.6 & $-48$:14:22 & $ 42.50\pm  1.89$ &  $<$29.31 &    7.5 &    0.22  &  2.54 &   1.64 & $< 1.43 $ & 17.71 &    (1.2e+09) &  PSPC & \multicolumn{2}{l}{dM2 in field}  \\
HD 48265 b          &        G5V & 06:40:01.7 & $-48$:32:31 & $ 87.40\pm  5.50$ &     29.53 &    6.0 &    0.16  &  1.16 &   1.51 & $  1.72 $ & 18.10 &      2.4e+09 &  PSPC & \multicolumn{2}{l}{}  \\
HD 70573 b          &    G1-1.5V & 08:22:50.0 & $+01$:51:33 & $ 45.7:$ &     29.09 &    4.0 &    0.30  &  6.10 &   1.76 & $  1.15 $ & 17.74 &      6.3e+08 &  PSPC & \multicolumn{2}{l}{uncertain $d$}  \\
HD 87883 b          &        K0V & 10:08:43.1 & $+34$:14:32 & $ 18.10\pm  0.31$ &     27.60 &    N/A &    2.73  &  1.78 &   3.60 & $ -0.96 $ & 17.07 &      4.9e+06 &  PSPC & \multicolumn{2}{l}{}  \\
HD 89744 b          &        F7V & 10:22:10.6 & $+41$:13:46 & $ 40.00\pm  1.06$ &     28.11 &    7.6 &    1.28  &  7.99 &   0.89 & $  0.76 $ & 18.78 &      2.6e+08 &  PSPC & \multicolumn{2}{l}{}  \\
HD 128311 b         &        K0V & 14:36:00.6 & $+09$:44:47 & $ 16.60\pm  0.27$ &     28.48 &    7.5 &    0.74  &  2.18 &   1.10 & $  0.95 $ & 18.01 &      4.0e+08 &  PSPC & \multicolumn{2}{l}{}  \\
HD 128311 c         &            &            &             &                   &           &        &          &  3.21 &   1.76 & $  0.54$ & 17.60  &      1.6e+08 & & &  \\
HD 142415 b         &        G1V & 15:57:40.8 & $-60$:12:00 & $ 34.20\pm  1.00$ &     28.65 &    5.0 &    0.57  &  1.62 &   1.05 & $  1.16 $ & 18.34 &      6.4e+08 &  PSPC & \multicolumn{2}{l}{}  \\
HD 147513 b         &     G3/G5V & 16:24:01.3 & $-39$:11:34 & $ 12.90\pm  0.14$ &     28.90 &   16.2 &    0.40  &  1.00 &   1.26 & $  1.25 $ & 18.14 &      8.0e+08 &  PSPC & \multicolumn{2}{l}{}  \\
HD 150706 b         &         G0 & 16:31:17.6 & $+79$:47:23 & $ 27.20\pm  0.42$ &     28.82 &   12.0 &    0.45  &  1.00 &   0.82 & $  1.54 $ & 18.52 &      1.6e+09 &  PSPC & \multicolumn{2}{l}{}  \\
HD 169830 b         &        F8V & 18:27:49.5 & $-29$:49:00 & $ 36.32\pm  1.20$ &     28.26 &   16.8 &    1.02  &  2.88 &   0.81 & $  0.99 $ & 18.80 &      4.4e+08 &  PSPC & \multicolumn{2}{l}{}  \\
HD 169830 c         &            &            &             &                   &           &        &          &  4.04 &   3.60 & $ -0.30$ & 17.50  &      2.2e+07 & & &  \\
HD 285968 b         &      M2.5V & 04:42:55.8 & $+18$:57:29 & $  9.40\pm  0.22$ &     27.41 &    3.2 &    3.62  &  0.03 &   0.07 & $  2.32 $ & 19.69 &      9.4e+09 &  PSPC & \multicolumn{2}{l}{}  \\
HR 810 b            &  G0V       & 02:42:33.5 & $-50$:48:01 & $ 15.50\pm  0.16$ &     28.79 &    7.0 &    0.47  &  1.94 &   0.91 & $  1.42 $ & 18.48 &      1.2e+09 &  PSPC & \multicolumn{2}{l}{}  \\
$\upsilon$ And b    &        F8V & 01:36:47.8 & $+41$:24:19 & $ 13.47\pm  0.13$ &  $<$28.24 &    6.5 &    1.06  &  0.69 &   0.06 & $< 3.25 $ & 21.02 &    (7.9e+10) &  PSPC & \multicolumn{2}{l}{K-M in field}  \\
$\upsilon$ And c    &            &            &             &                   &           &        &          &  1.92 &   0.83 & $< 0.95$ & 18.72  &    (4.0e+08) & & &  \\
$\upsilon$ And d    &            &            &             &                   &           &        &          &  4.13 &   2.51 & $<-0.01$ & 17.76  &    (4.4e+07) & & &  \\
%---------------------------------------------
\hline
\end{tabular}
\end{scriptsize}
\end{center}
\vspace{-3mm}
 \scriptsize{
 {\it Notes:} $^a$Planet data from The Extrasolar Planets
 Encyclopedia (http://exoplanet.eu). $^b$ XMM-Newton, Chandra or ROSAT instrument used for the X-ray flux. $^c$ 1~M$_{\rm J}$\,Gyr$^{-1}$=$6.02$e+13~g\,s$^{-1}$}
\end{table*}
}
%---------------------------------------------

%vvvvvvvvvvvvvvvvvvvvvvvvvvvvvvvvvvvvvvvvvvvvvvvvvvvvv
%\section{Results}\label{sec:results}

The stellar X-ray flux was
converted into X-ray luminosity, $L_{\rm X}$, using the distances 
listed in
Table~\ref{tabfluxes}. Also listed are some physical properties
of the hosted exoplanets, collected from the Exoplanets Database
(Schneider 1995, http://exoplanet.eu/).
The X-ray flux received at the orbit
of the planet is then given by $F_{\rm X}=L_{\rm X}/(4 \pi a_{\rm p}^2)$, where
$a_{\rm p}$ is the semimajor axis. 
The mass-loss rate from atmospheric losses with $\beta=K=1$
produced by X-rays simplifies to 
\begin{equation}
 \dot M_{\rm X} \sim \frac{3 F_{\rm X}}{{\rm G} \rho}
\end{equation}
Table~\ref{tabfluxes} includes this calculation for a
density of $\rho$=1.0\,g\,cm$^{-3}$. As reference
Jupiter, HD~209458b, and HD~189733b have $\rho$=1.24, 0.37
and 0.95 \,g\,cm$^{-3}$ respectively. 

The distribution of $F_{\rm X}$ against the planet mass,
$M_{\rm p} \sin i$, is plotted in Fig.~\ref{masses}. There is a separation
that seems to be related to mass. We plotted a line that
roughly follows this separation:  $\log F_{\rm X}=3 - 0.5 (M_{\rm p} \sin
i)$, with $M_{\rm p}$ in Jovian masses and $F_{\rm X}$ in CGS units.
This line is not based on any previous assumption or
physical law. 
We also include the Solar System planets in the diagram by using
current emission of the solar corona, 
which ranges $26\leq\log L_{\rm X}\leq27.7$ \citep{orl01},
with the vertical segments indicating the variations over the solar cycle. 
In this context we can compare the radiation arriving at the Earth
when life first appeared about
$\sim$3.5~Gyr ago \citep[see][and references therein]{cno07}
and at an 
earlier stage, to see whether coronal erosion could have
affected the Earth at that time. We use two stars considered
proxies of the Sun at an early age \citep{rib05}, $\kappa$~Cet
($\sim$1 Gyr, $\log L_{\rm X}=28.89$) and EK~Dra ($\sim$0.1 Gyr, $\log
L_{\rm X}=30.06$)\footnote{$L_{\rm X}$ calculated from XMM-Newton data}.
Lines indicating their $F_{\rm X}$ received at
1~a.u. 
are plotted in Fig.~\ref{masses} to mimic the flux at the Earth's orbit
in the past.

Since the effects of erosion accumulate over the planet's lifetime,
we alculated the integrated X-ray flux that has arrived on the
planet orbit between the
age of 20~Myr, when most protoplanetary disks would have dissipated, and the
present day.
We need to know the stellar age and the X-ray
luminosity evolution with time for each star in the sample. We can estimate
both following Garc\'es et al. (2010, in prep), who
relate the average X-ray luminosity to the age of late F to early M dwarfs:
{\setlength\arraycolsep{2pt}
\begin{eqnarray}
L_{\rm X} & = & 6.3\times10^{-4}\, L_{\rm bol}  \quad \qquad  (\tau < \tau_i)
\nonumber\\
L_{\rm X} & = & 1.89\times10^{28}\, \tau^{-1.55} \qquad  (\tau > \tau_i)
\end{eqnarray}}
with $\tau_i = 2.03\times 10^{20} L_{\rm bol}^{-0.65}$.  
$L_{\rm X}$ and $L_{\rm bol}$ are in erg\,s$^{-1}$, and $\tau$ is the
age in Gyr. The
relation was found with independent age indicators and/or wide binary
coeval companions to X-ray sources. The $\tau_i$ parameter marks the
typical change from saturation regime to an inverse proportionality
between $L_{\rm X}/L_{\rm bol}$ and rotation period
\citep[e.g.][]{piz03}. 
The calculation is taken as a
first approximation of the stellar age (Table~\ref{tabfluxes}),
considering also that there 
is an uncertainty of about an order of magnitude in the $L_{\rm X}$ levels
of stars of the same spectral type and age \citep{pen08b,pen08a}. The
accumulated X-ray flux 
at the planet orbit is shown in Fig.~\ref{agemasses}.
Subgiants are marked with different symbols since it is not known whether
they follow the same relation. 
A hard limit of $\sim 10^{22.14}$ erg\,cm$^{-2}$ in 10 Gyr is found by
combining the highest luminosity ($L_{\rm bol}=10^{34.5}~{\rm erg}~{\rm s}^{-1}$)
and the shortest distance 
to the star (0.02 a.u.) of planets in the sample. 
No higher values are
expected to be found in future observations of ``hot Jupiters''.
Our highest flux is $10^{21.23}$ erg\,cm$^{-2}$.

\begin{figure}[t]
   \centering
   \includegraphics[angle=90,width=0.49\textwidth,clip]{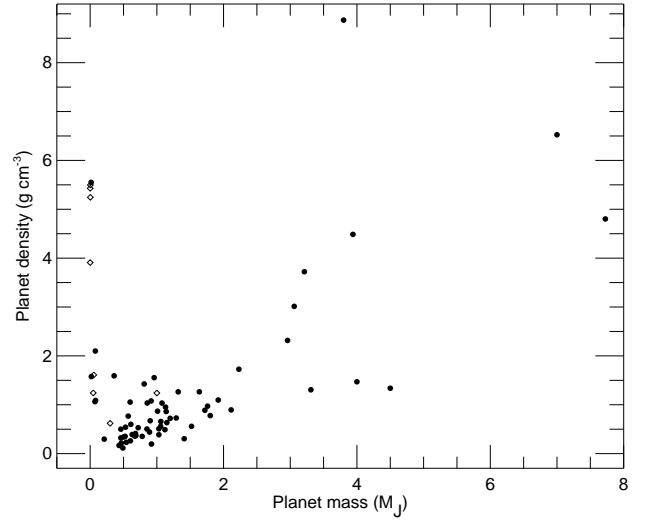}
   \caption{Density of the 66 planets of known radius (7 Jan. 2010) 
     with $M_{\rm p} < 8$\,M$_{\rm J}$ (filled circles). Diamonds
     represent the Solar System planets. 
   }\label{density}  
    \end{figure}

The effects of erosion in the long term are also expected to have
  an effect on the density of the population of close-in planets. The
  valuable information regarding the density is provided in most cases by
  the transit technique, that favours detection of planets with
  short periods, hence short distances. Our sample only has
  four planets with known density (HD 209458 b, HD 189733 b, Gj 436 b, and
  2M1207 b), but we can check the distribution of density with mass
  (Fig.~\ref{density}). This distribution is not representative of the
  whole population of exoplanets, and its results should only
  apply to close-in planets since erosion effects might be relevant.

%----------------------------------  Fig. 3
%----------------------------------------------

\section{Discussion}\label{sec:disc}
The observed sample seems to indicate an ``erosion
line'' (Fig.~\ref{masses})  
below which most planets are located. 
There are few planets above the erosion line, and they are probably at
an early evolutionary stage and have spent less time exposed to high
$F_{\rm X}$. 
The long-term accumulation effects are clearer in
Fig.~\ref{agemasses}, which shows that only 3 out of 34 planets above
1.5\,M$_{\rm J}$, have survived a flux of
$10^{19}$\,erg\,cm$^{-2}$, although the determination of this
  flux could be wrong for two of them (see below). This plot  
partially removes the effect of age.
Following dissipation of the protoplanetary disk,
planets exposed to high radiation should suffer heavy erosion,
until the X-ray flux
decreases as stellar rotation slows or the planet has become
small enough for gravity or
magnetospheric trapping to halt erosion. 
The thermal losses (Eq.~2) indicate that $F_{\rm X}$ and 
density control the mass loss rate. 
The dependence of the erosion line on mass, combined with the mass
distribution observed in Fig.~\ref{agemasses}, confirms that $F_{\rm X}$
is the main variable, with few massive planets surviving
exposure to high radiation as discussed below.
The distribution of density with mass displayed in Fig.~\ref{density}
is also consistent with the effects of erosion, since
planets with higher densities would suffer less erosion, resulting
in a population of massive planets in the long term that are denser than
lower mass planets. Gaseous planets should not
substantially increase their density, while being eroded above
jovian-like masses. Note also
that Eq.~2 is only valid for gaseous planets, and rocky planets
should suffer little erosion from XUV radiation.

In addition to thermal evaporation, non-thermal losses,
such as ion pick up and sputtering processes, could also be important
following ionization of the outer atmosphere
by the coronal radiation or high-energy particles
mediated by any planetary magnetic field.
Indeed, it is possible that the observations already reflect
the relative weakness of magnetic fields in massive
planets and the consequent inability to slow erosion.
Low-mass planets, with a wide range of
densities and distances in this sample, might have stronger
fields that reduce erosion. 

The planets $\tau$~Boo b, HD~195019 b and Gl 86 b,
seem to challenge this
interpretation (Fig.~\ref{agemasses}), retaining high masses despite
the high X-ray flux 
received. However, the fact that we see a young planet,
$\tau$~Boo b (age $\sim 400$ Myr, according to Eq. 3), still
suffering heavy erosion ($\dot M_{\rm X}$=0.011\,M$_{\rm J}$\,Gyr$^{-1}$ 
for $\rho$=1\,g\,cm$^{-3}$) reinforces
our interpretation of the erosion line. 
The age and accumulated X-ray flux 
determination of HD~195019, a G3IV-V star, could be
inaccurate, although it would have to be much younger for
the accumulated X-ray flux to be substantially reduced. This object
falls below the erosion line though. The third case, Gl 86, has a
white dwarf at 
only $\sim$21\,a.u. \citep{els01,mug05}, and although its
contribution to the X-ray 
flux should not be important, we cannot discard that dynamical
processes have changed the distance of the planet to the star over
its lifetime \citep[see also][]{lag06}. Finally, it
is possible that these three planets have higher densities that
have partly protected them against erosion.

No significant evaporation is currently underway in Solar System
planets, consistent with their location in Figs.~\ref{masses},
\ref{agemasses}.   
It is interesting to see that, when life
appeared on Earth, the planet was well below the ``erosion line'',
so probably suffering little or no erosion. Even 
100 Myr after the Sun was born, the Earth was still below the
the erosion line. By contrast,
any atmosphere on Mercury, much nearer the Sun (0.47 a.u.),
would have been stripped away.

%\subsection{Effects of biases}
The sample of known exoplanets is by no means complete.
Several selection effects could be present. ({\em i}) With the
methods used in exoplanet surveys, it is easier to
detect massive planets close to stars. 
This should bring more massive planets
with high $F_{\rm X}$ into our sample, thereby yielding a positive relation
of $F_{\rm X}$ with planet mass. This bias reinforces our
conclusions, since we find the opposite effect: the more massive
planets receive a lower $F_{\rm X}$. 
({\em ii}) Initial conditions in the disk could yield to
more massive planets placed at longer distances. The current sample
  of extrasolar planets has a deficit
  of massive planets at the shortest distances. We assume
  that initial conditions have little impact within this range of distances
  but we cannot exclude that the observed distribution is an effect of
  planet formation.
({\em iii}) X-ray luminous stars
are easier to detect in X-rays surveys, so we
should be biased towards planets with high $F_{\rm X}$ (very few
present in our sample), independent on planet mass.  In the other
side, the planet-hunting programs generally discard the active (young)
stars. Among these stars we would
likely find more planets above the ``erosion line'', but 
the planet would still be under heavy erosion for a young star, in good
agreement with our interpretation of the data. 
({\em iv}) Finally, most of the planets in the sample have $M_{\rm p}
\sin i < 2.5$\,M$_{\rm J}$, with few planets above this mass, 
that would actually be the most useful objects for confirming our
conclusions. 

Our approach is an alternative to the one followed by \citet{lec07}, who
balances the potential energy of the planet with the EUV flux received
from the star, based on a number of assumptions for estimating
the present and past EUV flux (with no estimation of the age of each
star), and the radius and composition of the planet. In particular,
the EUV flux is estimated using the flux in the range
110--200~\AA\ typical of each spectral type, and then scaled to the range
100--1200~\AA\ based on the 
solar pattern, an approach discouraged by the differences seen in the
few known EUV spectra in the literature \citep[see, e.g.,][]{san03b}.
\citet{lec07} extend this calculation for individual cases, most
notably claiming that Gj 876 b must be dense to have
survived the large estimated EUV flux. Our real measurements (close to
the ROSAT value of $\log L_{\rm X}=26.5$) indicate that this planet
receives less coronal radiation than others with low density, such
as HD 189733~b. \citet{dav09} instead balance the potential energy
with the X-ray flux that would arrive just during the saturation
period of stellar evolution, averaged according to spectral type. 
This would be a lower limit of the XUV flux, as the EUV band is
missing. As a result they suggest there is a
``destruction limit'' below which only dense planets would
survive. The present sample of exoplanets has at least three planets
(GJ 1214 b, GJ 436 b, HAT-P-12 b)
with low densities ($\rho=$1.58, 1.06, and 0.30
g\,cm$^{-3}$, respectively) below this limit.

\section{Conclusions}
Among 75 extrasolar planets, only 3 out of 34
high-mass planets of $M_{\rm p} \sin i>1.5$ (one of them still young) 
have been exposed to the
levels of radiation suffered by most of the low-mass planets in the
sample. 
We suggest that this is a consequence of the long-term effects of
erosion of gaseous planets by coronal radiation.
We propose the existence of an ``erosion line'' that depends on
the planet mass for the mass range explored, indicating that erosion
would have stronger effects for more massive planets. 
The heterogeneity of our sample makes it difficult to apply 
the thermal erosion models in the long-term, but these models cannot 
explain the observed dependence on mass. 
More complex models are required that consider
the chemical composition that could be very different among
planets: different molecules and ionization stages have different
responses to XUV radiation.
Non-thermal losses should also be considered, against which the
presence of a planetary magnetic field might provide protection.
Planets above the erosion line, such as the young $\tau$~Boo b, 
would be good candidates to search for
current effects of erosion by coronal radiation.
Finally we cannot exclude that the observed distribution is not
  partly an effect of planetary formation processes that would result in
  massive planets on wider orbits. More observations of X-ray
emission of planets with $M_{\rm p} \sin i \ga 2.5$\,M$_{\rm J}$ are
needed to confirm our conclusions.

\begin{acknowledgements}
      JSF and DGA acknowledge support from the
      Spanish MICINN through grant AYA2008-02038 
      and the Ram\'on y Cajal Program ref. RYC-2005-000549. 
      IR acknowledges support from the Spanish MICINN via grant
      AYA2006-15623-C02-01.
      This research has made use of the NASA's High Energy
      Astrophysics Science Archive Research Center (HEASARC) 
      and the public archives of XMM-Newton and Chandra.
      We thank
      G. Miniutti for help with the extragalactic source
      near 47 UMa. We are grateful to the anonymous referee and to the
      editor, T. Guillot, for the careful reading of and useful comments
      on the manuscript.
\end{acknowledgements}

\end{document}